\documentclass[prl,twocolumn,10pt,aps,superscriptaddress,longbibliography]{revtex4-1} 

\setcitestyle{square}
\usepackage{graphicx}  
\usepackage{float}
\usepackage{dcolumn}   
\usepackage{bm}        
\usepackage{amssymb}   
\usepackage{amsmath}
\usepackage{isomath}
\usepackage[usenames, dvipsnames]{color}
\usepackage{xcolor}
\usepackage{soul}
\usepackage{physics}
\usepackage{textcomp}
\usepackage[normalem]{ulem}
\usepackage{gensymb}
\usepackage{flushend}

\usepackage{hyperref}
\hypersetup{
     colorlinks  = true,
     linkcolor    = red,
     citecolor    = cyan,
     urlcolor     = magenta
     }

\begin{document}

\widetext

\title{Corner states in second-order mechanical topological insulator}

\author{Chun-Wei Chen}
\affiliation{Aeronautics and Astronautics, University of Washington, Seattle, WA 98195-2400, USA}
\author{Rajesh Chaunsali}
\affiliation{LAUM, CNRS, Le Mans Université, Avenue Olivier Messiaen, 72085 Le Mans, France}
\author{Johan Christensen}
\affiliation{Department of Physics, Universidad Carlos III de Madrid, ES-28916 Legan\`es, Madrid, Spain}
\author{Georgios Theocharis}
\affiliation{LAUM, CNRS, Le Mans Université, Avenue Olivier Messiaen, 72085 Le Mans, France}
\author{Jinkyu Yang}
\email[Corresponding author.\\]{jkyang@aa.washington.edu}
\affiliation{Aeronautics and Astronautics, University of Washington, Seattle, WA 98195-2400, USA}
\date{\today}

\begin{abstract}
We numerically and experimentally study corner states in a continuous elastic plate with embedded bolts in a hexagonal pattern. While preserving $C_6$ crystalline symmetry, the system can transition from a topologically trivial to a non-trivial configuration. We create interfacial corners of 60$\degree$ and 120$\degree$ by adjoining trivial and non-trivial topological configurations. Due to the rich interaction between the bolts and the continuous elastic plate, we find a variety of corner states with and without topological origin. Notably, some of the corner states are highly localized and tunable. 
By taking advantage of this property, we experimentally demonstrate one-way corner localization in a Z-shaped domain wall. 

\end{abstract}

\keywords{}
\maketitle

\textit{Introduction}.---Topological insulators provide researchers with efficient ways to tailor and control the energy flow. These topologically non-trivial phases have drawn growing attentions, since the immunity to back-scattering -- the key feature of topological protection -- can help overcome defects and sharp bends during energy transfer. 
Realization of these topological boundary states have been demonstrated in classical systems, such as acoustics and mechanics, through mimicking the quantum Hall effect \cite{nash2015topological,wang2015topological}, the quantum spin Hall effect \cite{mousavi2015topologically,he2016acoustic} or the quantum valley Hall effect \cite{lu2017observation, pal2017edge, Liu2018valley}. Recently, higher-order topological insulators with multiple moments have been predicted theoretically \cite{benalcazar2017quantized, benalcazar2017electric, schindler2018higher}, and parallel experimental works of the second-order topological quadrupole insulators have been demonstrated in mechanics \cite{serra2018observation}, microwave circuits \cite{peterson2018quantized}, electrical circuits \cite{imhof2018topolectrical}, photonics \cite{mittal2019photonic}, and acoustics \cite{Qi2020quad}. 

To create a topological quadrupole insulator, negative hopping parameter is a requisite ingredient. 
In practice, however, the negative coupling needs much effort to design. 
Therefore, another way based on \textit{bulk} dipole moments has been proposed to form a second-order topological insulator. By leveraging the protection of crystalline symmetry, it shows the in-gap corner states due to the ``filling anomaly'' \cite{benalcazar2019quantization}. The corner states of such second-order topological crystalline insulators with vanishing quadrupole moment have been studied primarily in square \cite{xie2018second, chen2019direct, chen2019corner, zhang2019deep, zhang2019second,zhang2019non, lopez2019multiple, Coutant2020SSH} and Kagome lattices \cite{xue2019acoustic, ni2019observation, el2019corner, chen2019effect, kempkes2019robust}. 
Recent studies in photonics \cite{Jiho2018photonics, xie2020higher} have shown that hexagonal lattices can also support corner modes with interesting properties, but their mechanical analogue has been limited to \textit{discrete} mechanical structures which lack the engineering potential and practicality \cite{fan2019elastic}. 

In this Letter, we propose a bolted plate structure as a platform of a \textit{continuous} second-order mechanical topological insulator. The plate is decorated with bolts, which act as resonators, arranged in $C_6$-symmetric hexagonal lattice. For the topological characterization, we approximate our system with a lumped-mass model \cite{Torrent2013plate}. 
Then, the topological indices are determined based on the rotational symmetries of eigenmodes at the high-symmetry points in the Brillouin zone (BZ). 
By joining two topologically-distinct configurations, we report the generation of two different types of corner modes: one with topological origin and the other without it. Interestingly, we find that the one without topological origin exhibits higher localization and tunability than the topological one. By leveraging these characteristics, we experimentally demonstrate a one-way corner localization of mechanical waves in a Z-shaped domain wall. This asymmetric wave localization mechanism can be used for advanced control of energy flow.


\begin{figure}[!]
\centering
\includegraphics[width=\columnwidth]{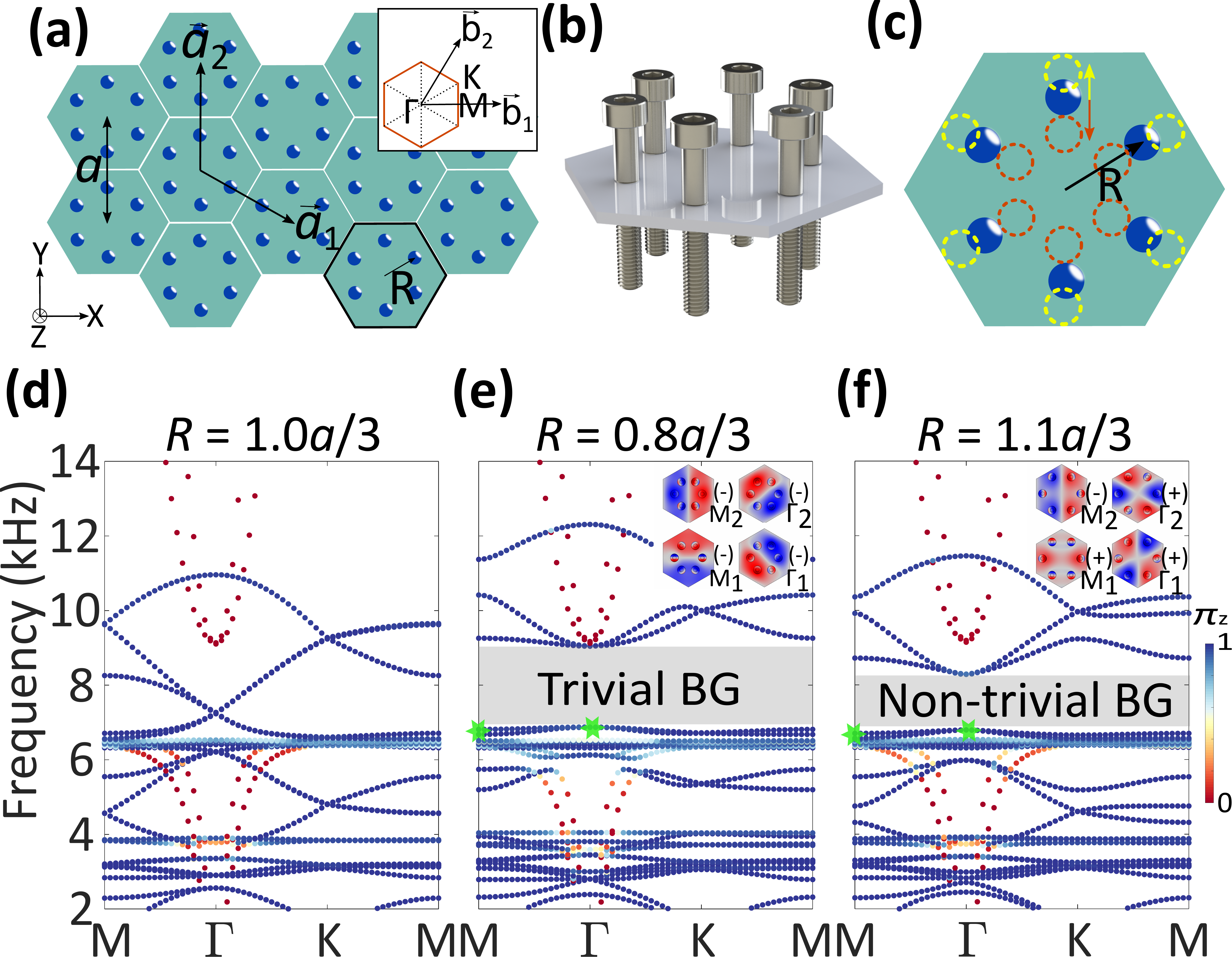}
\caption{(a) Continuous plate structure with hexagonally arranged bolts (blue dots). The translation vectors $\vec{a}_{1}$ and $\vec{a}_{2}$ and the corresponding reciprocal vectors $\vec{b_1}$ and $\vec{b_2}$ in the first Brillouin zone (inset). (b) A graphical illustration of the unit cell. (c) An enlarged plot of the unit cell with expanded ($R > a/3$, yellow) and shrunk ($R < a/3$, red) arrangements of the mounted bolts. (d) $R=a/3$ case, which is exactly the honeycomb lattice with a double Dirac cone at the $\Gamma$ point. Colorbar represents the level of the out-of-plane motion. (e) $R=0.8a/3$ case, which leads to the emergence of a trivial band gap represented in grey rectangle. Inset shows the four corresponding eigenmodes extracted from the marked green stars at the $M$ and $\Gamma$ points. (f) $R=1.1a/3$ case with a non-trivial band gap.}
\label{fig1_corner}
\end{figure} 

\textit{System and unit-cell dispersion}.---
In Fig.~\ref{fig1_corner}(a), steel bolts are mounted hexagonally in an aluminum plate. See Fig.~\ref{fig1_corner}(b) for the graphical illustration of the unit cell. The lattice constant is $a=45$ mm.
\textit{R} is the circumferential radius of six bolts, which is a tuning parameter of creating the trivial and non-trivial band gap. 
The enlarged unit cell in Fig.~\ref{fig1_corner}(c) shows that as \textit{R} increases, the six bolts gradually expand until reaching the limit of the unit cell's boundary as represented in yellow; whereas, as the \textit{R} decreases, the six bolts gradually merge into the center as indicated in red. 
The band structure of $R=1.0a/3$, $R=0.8a/3$, and $R=1.1a/3$ are shown in Figs.~\ref{fig1_corner}(d), \ref{fig1_corner}(e), and \ref{fig1_corner}(f), respectively, based on finite element analysis (FEA, see Supplemental Material~\cite{SM} for more details). The color bar quantifies the dominance of plate displacement in out-of-plane ($z$) direction, which is defined as $\Pi_z= \frac {\int_{V} \mid w \mid ^2dV}{\int_{V} (\mid u \mid ^2 + \mid v \mid ^2 + \mid w \mid ^2)dV}$, where $V$ is the volume of the plate of a unit cell and $u$, $v$ and $w$ are the displacement components in $x$, $y$ and $z$ axes. When $\Pi_z = 1 $, it means that the eigenmode is completely dominated by the out-of-plane motion; whereas, when $\Pi_z = 0 $, the eigenmode is entirely dominated by the in-plane motion. From Fig.~\ref{fig1_corner}(d), we see that there is a double Dirac cone at the $\Gamma$ point at 7.27 kHz. Once the radius $R=a/3$ is no longer maintained, the double Dirac point opens and creates a band gap [Figs.~\ref{fig1_corner}(e) and \ref{fig1_corner}(f)]. 

\textit{Topological characterization}.---
We perform the topological characterization of the band gaps based on the pseudospins of eigenmodes at the $\Gamma$ point~\cite{chaunsali2018experimental}. This analogue of the quantum spin-Hall effect helps us predict the existence of chiral \textit{edge} states at the interface between two domains made of shrunk ($R < a/3$) and expanded ($R > a/3$) unit cells. In the present study, however, we are interested in the \textit{corner} states, and therefore, we characterize the band gaps based on quadrupole \cite{liu2019helical} or rotation invariants \cite{Jiho2018photonics, benalcazar2019quantization}. These rely on the parity (the eigenvalue of $\pi$ rotation over the $z$ axis) of eigenmodes at the $\Gamma$ and $M$ points of the BZ for \textit{every} band below the band gap.
In the insets of Figs.~\ref{fig1_corner}(e) and \ref{fig1_corner}(f), we plot eigenmodes for the first \textit{two} bands immediately below the band gap at the $\Gamma$ and $M$ points (light green stars) and calculate their parity.
For the shrunk configuration, the two bands have $-1$ parity at the $\Gamma$ and $M$ points. However, for the expanded configuration, the two bands have +1 parity at the $\Gamma$ point, but the opposite parity at the $M$ point.

Ideally, this characterization process needs to be repeated for all bands below the band gaps. This is a complicated task due to the existence of the numerous dispersion curves [see Figs.~\ref{fig1_corner}(e) and \ref{fig1_corner}(f)], which result from the multi degrees of freedom and coupling between the bolt and the plate. To simplify this, we approximate the system into a lumped-mass model, in which the bolts are modeled as point masses connected to the plate with transverse springs \cite{Torrent2013plate} (see Supplemental Material \cite{SM} for more details on this model and its validity). 
As a result, we can calculate the parity for all the bands below the band gap, and obtain quadrupole 0 and 1/2, and rotation invariants $[0,0]$ and $[2,0]$ for the shrunk (trivial) and expanded (nontrivial) unit cells, respectively. This establishes topological distinction of the two band gaps shown in Figs.~\ref{fig1_corner}(e) and \ref{fig1_corner}(f). 

\begin{figure}[!]
\centering
\includegraphics[width=\columnwidth]{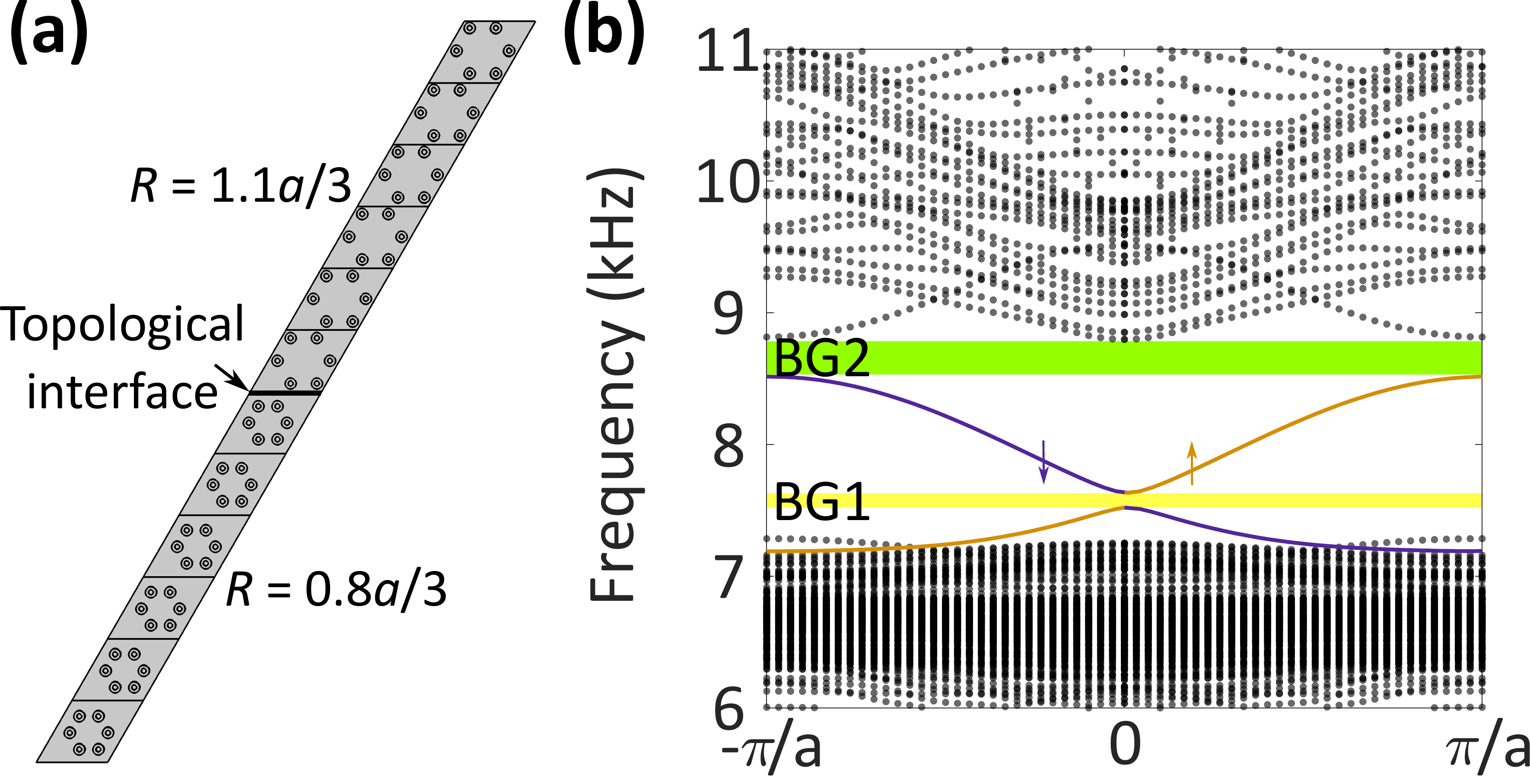}
\caption{(a) A supercell made by placing six non-trivial ($R=1.1a/3$) and six trivial ($R=0.8a/3$) cells adjacently. (b) Eigenfrequencies of the supercell as a function of wave numbers in the periodic direction. Bulk bands are in black. There are two edge modes with opposite pseudospins (purple and yellow) inside the bulk band gap, where two mini gaps (BG1 and BG2) are generated.}
\label{fig2_corner}
\end{figure} 

\textit{Supercell analysis}.---
We return to the FEA approach and first perform a supercell analysis by adjoining six non-trivial ($R=1.1a/3$) and six trivial cells ($R=0.8a/3$) [Fig.~\ref{fig2_corner}(a)]. The terminations are free, and the sides are Floquet-periodic. We observe that the interface hosts chiral propagating modes inside the bulk band gap of the supercell dispersion in Fig.~\ref{fig2_corner}(b). These propagate in opposite directions with two opposite pseudospins (clockwise and counterclockwise in purple and yellow, respectively), mimicking spin Hall effect \cite{Wu2015spinHall}. However, these edge states are gapped as opposed to the quantum spin-Hall effect for fermions that supports \textit{gapless} edge states protected by the time-reversal symmetry \cite{Kane2005SpinHall}. Even though there are ways to close these gaps for a broadband wave propagation at the interface \cite{liu2019nonconventional}, we deliberately use them to look for potential corner modes in this study. BG1 is at low frequency and resides \textit{within} the interface mode spectrum. This emerges due to the breakage of crystalline $C_6$ symmetry at the interface and exists as long as there are cells across the interface with different \textit{R} (see the details in Supplemental Material \cite{SM}). BG2 is at a higher frequency and lies \textit{above} this spectrum and below the bulk spectrum (black) and emerges due to the \textit{large} mismatch of radii between trivial and nontrivial bolted-plate lattices (Supplemental Material \cite{SM}). 

\begin{figure}[!]
\centering
\includegraphics[width=\columnwidth]{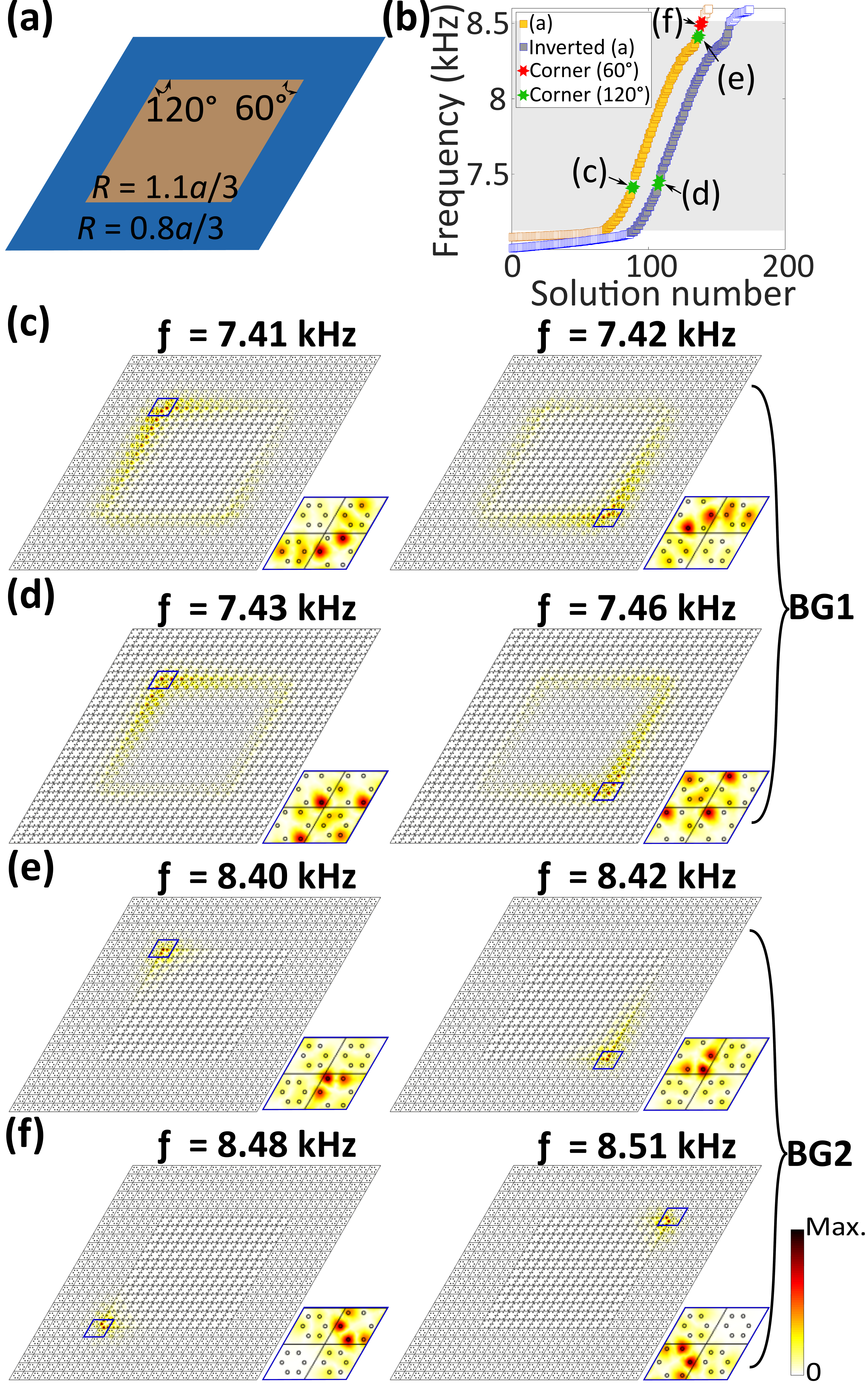}
\caption{(a) A rhombus-shaped structure with an interface between two domains: $R=1.1a/3$ (inner) and $R=0.8a/3$ (outer). (b) The eigenfrequency for the configuration in (a) and its inverted counterpart. Bulk band gap is marked in grey. Green and red stars represent the corner states. (c)--(f) Eigenmodes of the corner states corresponding to the stars in (b). The color map represents the amplitude of the out-of-plane displacements, $\abs{w}$. }
\label{fig3_corner}
\end{figure} 

\textit{Emergence of corner states}.---To observe these corner states and investigate their differences systematically, we construct a rhombus-shaped, topologically \textit{non-trivial} domain ($R=1.1a/3$) inside the \textit{trivial} domain ($R=0.8a/3$) [Fig.~\ref{fig3_corner}(a)]. This contains two 120$\degree$ corners and two 60$\degree$ corners. We also consider the ``inverted'' configuration, in which the two domains are interchanged.
We then perform the eigenfrequency analysis on both configurations and show the results in Figs.~\ref{fig3_corner}(b). We observe the emergence of several corner states marked with the green and red stars. 

In Fig.~\ref{fig3_corner}(c), we show the low-frequency corner states for the rhombus-shaped structure of Fig.~\ref{fig3_corner}(a) while in Fig.~\ref{fig3_corner}(d) the corner modes of the inverted configuration. These corner states reside in the BG1. 
Importantly, we find that only 120$\degree$ corners support corner states. These states are of topological origin. For the verification, we parameterize our system with the unit cells with varying radii and find that these corner states exist robustly even for a small difference in radii between the trivial and nontrivial cells (see Supplemental Material \cite{SM} for details). Moreover, a corner state of the topological origin should persist when two domains are interchanged, since their existence is based on the \textit{difference} in the topological nature of the two adjacent domains. This is exactly what we observe in Figs.~\ref{fig3_corner}(c) and \ref{fig3_corner}(d). Upon closely examining the mode shapes of these states in Fig.~\ref{fig3_corner}(c), we find that the peak displacement within the non-trivial unit cell, located at the corner, occurs at the two bolts adjacent to the most cornered bolt, which shows the minimum displacement. This is consistent with the shapes of the topologically-nontrivial modes reported in photonics recently \cite{Jiho2018photonics}. The inversion of domains in Fig.~\ref{fig3_corner}(d), however, changes the mode shape, and now the peak displacement within the non-trivial unit cell occurs at the most cornered bolt.   

Interestingly, there are also high-frequency corner states, marked by red and green star in Fig.~\ref{fig3_corner}(b), which reside within BG2. We observe that \textit{both} 60$\degree$ and 120$\degree$ corners support these states as shown in Figs.~\ref{fig3_corner}(e) and \ref{fig3_corner}(f). However, these exist only in the regular configuration shown in Fig.~\ref{fig3_corner}(a), but not in the inverted configuration. This hints at their nontopological origin, which we verify by performing a parametric study again with varying radii of unit cells (see Supplemental Material \cite{SM} for details). We find that these states exist only for a large difference in radii between the trivial and non-trivial unit cells and are \textit{not} predicted by our simplified lumped-mass model, as opposed to the corner states within BG1. This suggests that they appear due to the complex interaction of bolt-plate assembly that the lumped-mass model fails to capture. Their mode shapes also differ from the ones in BG1. For example, the states shown in Fig.~\ref{fig3_corner}(e) have the peak displacement occurring at the most cornered bolt within the non-trivial unit cell, while the adjacent two bolts have also nonzero displacements. This is also consistent with the topologically-trivial corner modes reported in Ref.\cite{Jiho2018photonics}.

What makes the corner states observed in BG2 unique is their tunability when domains are interchanged, and also their spatial localization which is much higher compared to the corner states in BG1 [compare the localization lengths in Figs.~\ref{fig3_corner}(e) and \ref{fig3_corner}(f) to those in Figs.~\ref{fig3_corner}(c) and \ref{fig3_corner}(d)]. These characteristics will be leveraged to achieve one-way energy localization as will be described below. 

\begin{figure}[!]
\centering
\includegraphics[width=\columnwidth]{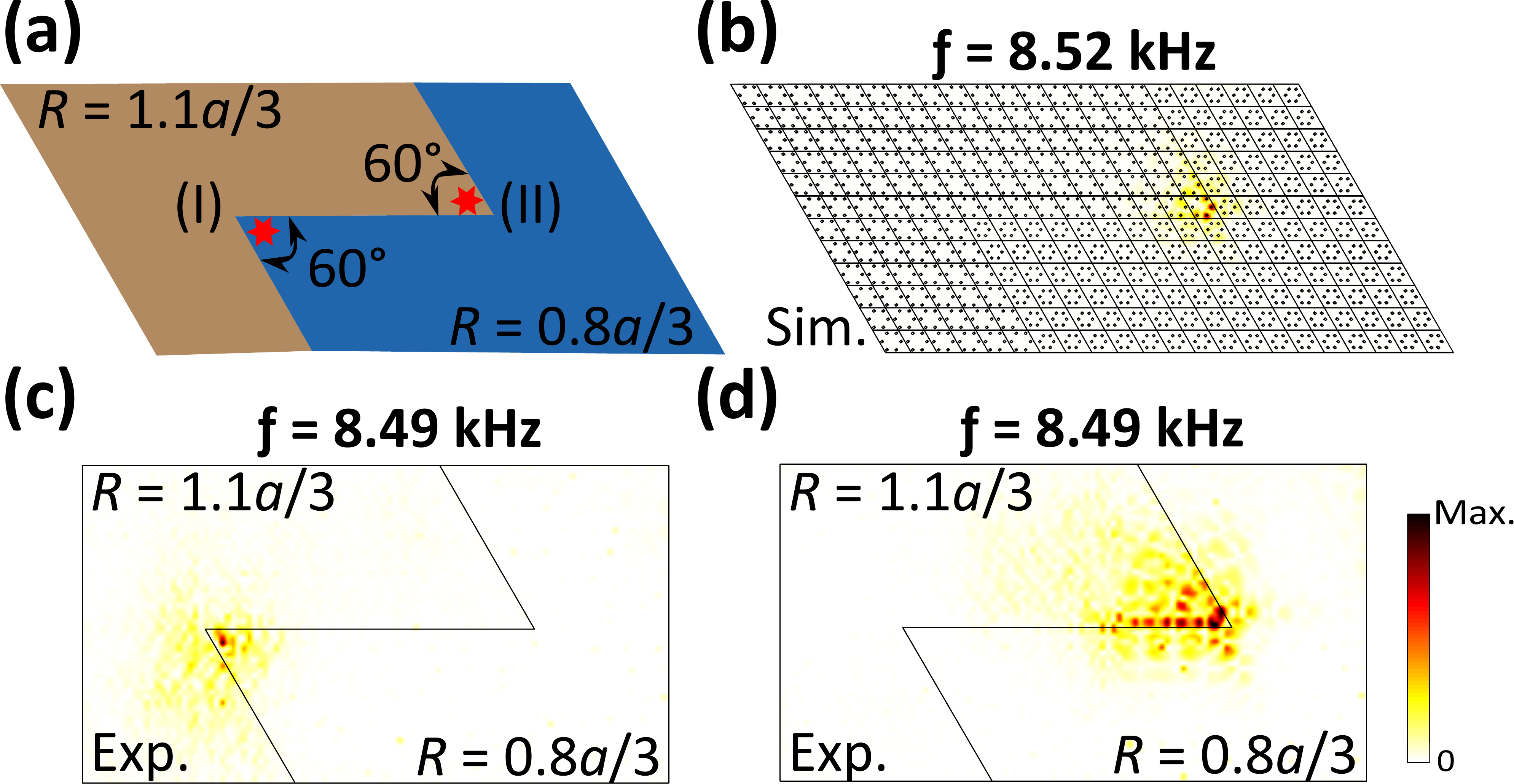}
\caption{Experimental verification of the corner state at 60$\degree$ corners. (a) A Z-shaped interface with two different 60$\degree$ corners is created via placing trivial cells ($R=0.8a/3$) with non-trivial cells ($R=1.1a/3$) adjacently. The red stars mark the locations of the piezo-actuators that excite the elastic plate. (b) A simulated eigenmode shows that corner state appear only at the corner (II) at $f=8.52$ kHz. (c)-(d) The measured wave-field response of the bolted plate when the piezo-actuator is attached on the corner (I) and corner (II), respectively, and excited at $f=8.49$ kHz.}
\label{fig4_corner}
\end{figure} 

\textit{Experimental demonstration}.---We build a Z-shaped domain wall that includes two different types of 60$\degree$ corners in one setup (Fig.~\ref{fig4_corner}(a), see Supplemental Material \cite{SM} for the fabrication and measurements detail). From the simulation results shown earlier, we know that the corner state without the topological origin exists only in the case when the non-trivial cells are surrounded by the trivial cells [Figs.~\ref{fig3_corner}(e) and \ref{fig3_corner}(f)], i.e., at corner (II) in Fig.~\ref{fig4_corner}(a). This is again verified by the eigen-frequency analysis on this particular setup. Fig.~\ref{fig4_corner}(b) shows a highly-localized corner state with $f=8.52$ kHz at the corner (II). 

We excite the plate using a piezoelectric ceramic actuator by placing it at corner (I) and (II) in two separate experiments. 
We use a chirp signal with the frequency range of 2--40 kHz. A point-by-point measurement is then conducted by using the laser Doppler vibrometer to detect the flexural waves. By gathering and reconstructing measured data from all the points, we plot the steady state wave-field at $f = 8.49$ kHz (Figs.~\ref{fig4_corner}(c, d)). When corner (I) is excited, there is no evidence of a corner state apart from the usual exponentially decaying evanescent field [Fig.~\ref{fig4_corner}(c)]. When we excite corner (II), however, we observe clear confinement of energy due to the presence of the corner mode [Fig.~\ref{fig4_corner}(d)]. In addition, the profile of the corner mode matches closely with the simulation results in that the last two resonators of the non-trivial unit cell have peak displacements [compare Fig.~\ref{fig4_corner}(b) and Fig.~\ref{fig4_corner}(d)]. It should also be noted that the frequencies of the corner state between the experimental and computational results are in excellent agreement. 

\begin{figure}[b!]
\centering
\includegraphics[width=\columnwidth]{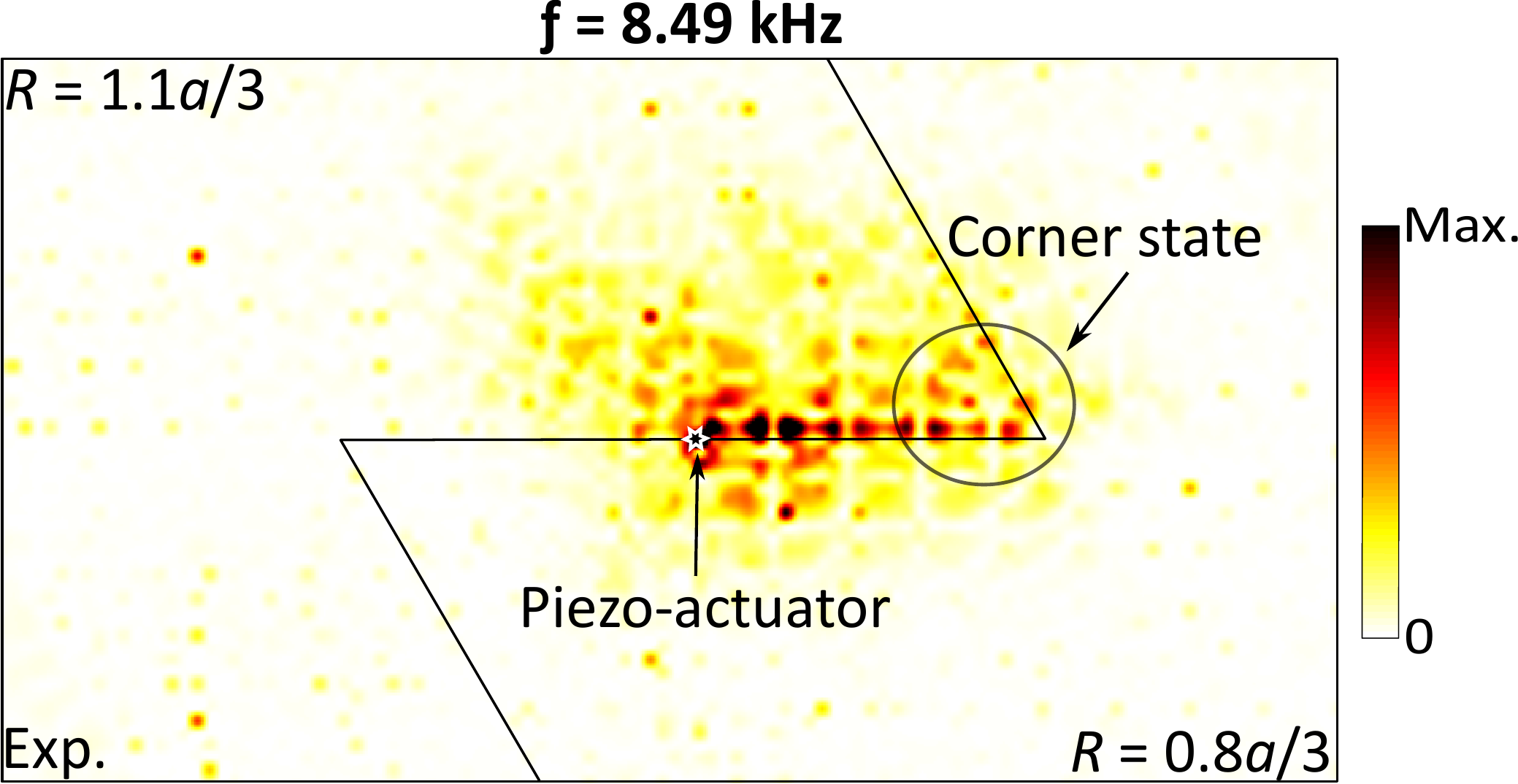}
\caption{Experimental demonstration of asymmetric wave localization when the piezo-actuator is attached in the middle of the Z-shaped interface.}
\label{fig5_corner}
\end{figure} 

Next, we exploit this selective localization observed in the previous test to demonstrate a one-way localization through the Z-shaped interface. We excite the plate with a harmonic excitation at $f=8.49$ kHz in the middle of the interface as shown in in Fig.~\ref{fig5_corner}. We observe that the flexural waves departing from the point of excitation propagate towards the right direction only, thereby exciting the corner (II) only, whereas the corner (I) being at the same distance does not see any such energy localization. Such asymmetric wave localization is a highly useful -- yet relatively unexplored -- feature that can be exploited to manipulate energy flow at will. 

\textit{Conclusions}.---We propose a ubiquitous design of a bolted plate in the hexagonal arrangement to demonstrate in-gap corner states in our $C_6$-symmetry-protected system. 
By changing the radius of the unit cell, we construct two configurations that show topologically distinct band gaps. We perform topological characterization of the bolted-plate assembly based on a simple lumped-mass model. When two such topologically distinct bolted-plates are placed adjacently, we conduct full geometry simulations to show that there are two regions (mini gaps) in frequency where different types of corner modes can exist. We find that the low-frequency corner states are of topological origin and the high-frequency corner states are of nontopological origin. While the former can be predicted by the lumped-mass model, the later can not. However, the later are highly localized for at \textit{both}  60$\degree$ and 120$\degree$ corners and can be made to exist or non-exist based on the inversion of topologically-distinct domains across the interface.
This fact is thus used to create a Z-shaped interface between topologically distinct domains for achieving an asymmetric localization of energy. We expect that these findings will enrich the wave-localization phenomena in mechanics and encourage new applications in vibration management.

\bigskip
C.-W. C. and J. Y. are grateful for the support from NSF (CAREER1553202 and EFRI-1741685). R. C. and G. T.  acknowledge support by the project  CS.MICRO funded under the program Etoiles Montantes of the Region Pays de la Loire. J. C. acknowledges the support from the European Research Council (ERC) through the Starting Grant 714577 PHONOMETA and from the MINECO through a Ram\'on y Cajal grant (Grant No. RYC-2015-17156). 

\bigskip
C.-W. C. and R. C. contributed equally to this work.

\def\bibsection{\section*{References}} 
\bibliographystyle{ieeetr}
\bibliography{Ref}

\end{document}